\documentclass[fleqn,usenatbib,useAMS]{mnras}

\usepackage{newtxtext,newtxmath}

\usepackage[T1]{fontenc}

\DeclareRobustCommand{\VAN}[3]{#2}
\let\VANthebibliography\thebibliography
\def\thebibliography{\DeclareRobustCommand{\VAN}[3]{##3}\VANthebibliography}

\usepackage{graphicx}	
\usepackage{amsmath}	
\usepackage{orcidlink}

\usepackage{xspace}
\usepackage{subcaption}

\newcommand{\phonenumber}{WD J0135\xspace}
\newcommand{\msun}{{\rm M}_{\sun}}
\newcommand{\rout}{r_{\mathrm{out}}}
\newcommand{\rwd}{r_{\mathrm{wd}}}
\newcommand{\mwd}{m_{\mathrm{wd}}}

\title[UMWD seismology and magnetism]{Seismology and diffusion of ultramassive white dwarf magnetic fields}

\author[D. Blatman et al.]{
Daniel Blatman$^{\orcidlink{0009-0001-1957-8801}}$,$^{1}$\thanks{E-mail: \href{mailto:daniel.blatman1@mail.huji.ac.il}{daniel.blatman1@mail.huji.ac.il} (DB);
\href{mailto:nrui@caltech.edu}
{nrui@caltech.edu} (NZR)
}\label{emails}\thanks{These authors contributed equally to this work.}\label{contrib}
Nicholas Z. Rui$^{\orcidlink{0000-0002-1884-3992}}$,$^{2}$\hyperref[emails]{\footnotemark[1]}\hyperref[contrib]{\footnotemark[2]}
Sivan Ginzburg$^{\orcidlink{0000-0002-3751-4553}}$$^{1}$
and Jim Fuller$^{\orcidlink{0000-0002-4544-0750}}$$^{2}$
\\
$^{1}$Racah Institute of Physics, The Hebrew University, Jerusalem 9190401, Israel\\
$^{2}$TAPIR, California Institute of Technology, Pasadena, CA 91125, USA
}

\date{Accepted XXX. Received YYY; in original form ZZZ}

\pubyear{\the\year{}}

\begin{document}
\label{firstpage}
\pagerange{\pageref{firstpage}--\pageref{lastpage}}
\maketitle

\begin{abstract}
Ultramassive white dwarfs (UMWDs; defined by masses $\gtrsim 1.1\,\msun$) are prime targets for seismology, because they pass through the ZZ Ceti instability strip at the same time that their cores crystallize.
Recent studies suggest that crystallization may magnetize white dwarf interiors with a strong magnetic field $B_0$ up to a radius $r_{\rm out}^0$, either through a magnetic dynamo or by transporting a pre-existing fossil field.
We demonstrate that seismology can probe these buried fields before they break out at the surface, because even the weak exponential tail of the outwardly diffusing field can disrupt the propagation of gravity waves near the surface.
Based on the observed oscillation modes of WD J0135+5722 -- the richest pulsating UMWD to date -- we constrain its surface field $B_{\rm surf}\lesssim 2\,\textrm{kG}$.
We solve the induction equation and translate this to an upper limit on the internal field $B_0$.
For a carbon--oxygen (CO) core we find $B_{\rm surf}\ll B_0\lesssim 0.6\,\textrm{MG}$, consistent with the crystallization dynamo theory. For an oxygen--neon (ONe) core, on the the other hand, $r_{\rm out}^0$ is larger, such that the magnetic field breaks out and $B_{\rm surf}\lesssim B_0\lesssim 7\,\textrm{kG}$.
This low magnetic field rules out an ONe composition or, alternatively, an intense dynamo during crystallization or merger.
Either way, the imprint of magnetic fields on UMWD seismology may reveal the uncertain composition and formation paths of these stars. 
\end{abstract}




\section{Introduction}

Observationally, white dwarfs are now known to possess a wide range of magnetic fields from $\lesssim10^3\,\mathrm{G}$ to $\sim10^9\,\mathrm{G}$ \citep[see][for reviews]{Ferrario2015,Ferrario2020}.
Although the origin of white dwarf magnetism has remained an open question for half a century \citep{Kemp1970,AngelLandstreet71,LandstreetAngel71}, recent volume-limited surveys show that the magnetic fields of typical-mass white dwarfs ($\lesssim0.8\,\msun$) tend to emerge late, only after several Gyr of cooling \citep{BagnuloLandstreet2022}.
These magnetic fields are either buried fossil remnants from previous stages of stellar evolution that diffuse to the surface \citep{Camisassa2024} or newly formed fields generated by a dynamo during the white dwarf cooling phase.
Specifically, core crystallization may play an important role, either actively by driving a magnetic dynamo \citep{Isern2017,Schreiber2021Nat,Ginzburg2022,Fuentes2023,Fuentes2024,BlatmanGinzburg2024,CastroTapia2024a,CastroTapia2024b,MontgomeryDunlap2024} or passively by transporting fields closer to the surface through convection.

Ultramassive white dwarfs (UMWDs; defined by masses $\gtrsim 1.1\,\msun$) are yet more mysterious. In addition to the unknown origin of their magnetic fields, UMWDs can harbour either carbon--oxygen (CO) or oxygen--neon (ONe) cores, in contrast to normal-mass white dwarfs, which are universally of CO composition. 
The actual UMWD population probably contains some of both: \textit{Gaia} observations of white dwarf cooling delays \citep{Cheng2019,Tremblay2019,Bauer2020,Blouin2021,Camisassa2021,Bedard2024} and measured abundances in novae \citep{TruranLivio1986,LivioTruran1994} provide evidence for the existence of CO and ONe UMWDs, respectively.
UMWDs' uncertain compositions are tied to their highly uncertain (and probably diverse) formation mechanisms.
In particular, UMWDs can be produced either by single stellar evolution or through the merger of two lower-mass white dwarfs.
In both cases, whether a CO or an ONe UMWD forms hinges on whether the system ever successfully ignites carbon, which in turn depends very sensitively on modelling details \citep{Dominguez1996,Althaus2021,Schwab2021,Wu2022,Shen2023}. 

\citet{BagnuloLandstreet2022} found that UMWDs appear to develop magnetic fields at younger cooling ages and with higher frequency than regular white dwarfs.
They interpreted this as a smoking gun for a dynamo operating during a double white dwarf merger which formed them \citep{GarciaBerro2012,Ji2013,Zhu2015}.
\citet{BlatmanGinzburg2024b}, on the other hand, demonstrated that a crystallization dynamo may also operate early enough at such high masses to explain at least part of the observations, depending on the white dwarf's composition.
Similarly, \citet{Camisassa2022} proposed that magnetic fields could be used to distinguish between CO and ONe UMWDs by exploiting the difference in their crystallization times.

Seismology is another powerful technique to probe white dwarf magnetism, because white dwarf pulsations are sensitive to relatively weak magnetic fields \citep[e.g.][]{Jones1989,Winget1994,Montgomery2010}. Recently, \citet{rui2025supersensitive} demonstrated that such pulsations could be used to place upper bounds on white dwarf surface magnetic fields that are well below the detection limit of spectroscopy and polarimetry \citep{Landstreet2015}.
The recent discovery of the partially crystallized ultramassive WD J0135+5722 \citep[hereafter \phonenumber;][]{DeGeronimo2025} opens the door to extending this analysis to UMWDs.
With 19 measured pulsation modes, \phonenumber is the richest pulsating UMWD known to date. 

Here, following \citet{rui2025supersensitive}, we place a seismic upper limit on the surface magnetic field of \phonenumber.
Then, similarly to \citet{CastroTapia2024b}, we solve the induction equation to relate the surface and core field strengths.
Because the crystallization dynamo is probably most efficient during the initial stages of crystallization \citep{CastroTapia2024a,Fuentes2024}, we assume that the magnetic field begins diffusing from the outer edge of the convection zone that forms during the onset of crystallization \citep{Isern2017,Camisassa2022,Bauer2023,BlatmanGinzburg2024}.
We demonstrate that this technique can probe buried magnetic fields before they fully break out to the surface, potentially revealing the formation channel, composition, and magnetic field origin of UMWDs.

The remainder of this paper is organized as follows.
In Section \ref{sec:models} we describe our numerical UMWD models and how we determine the outer radius of convection.
In Section \ref{sec:diffusion} we compute the magnetic field's diffusion from this outer radius to the surface.
In Section \ref{sec:seismology} we use seismology to place upper limits on the surface and internal magnetic fields of \phonenumber.
We summarize our conclusions in Section \ref{sec:conclusions}.

\section{White dwarf models}\label{sec:models}

We used the stellar evolution code \textsc{mesa} \citep{Paxton2011,Paxton2013,Paxton2015,Paxton2018,Paxton2019,Jermyn2023}, version r24.08.1, to create two white dwarf models with a CO and an ONe composition, both with a mass $\mwd=1.14\,\msun$, consistent with \phonenumber\,\citep{DeGeronimo2025}.
The ONe white dwarf was created by evolving a pre-main-sequence progenitor star through the different stages of stellar evolution using the \texttt{make\_o\_ne\_wd} test suite.\footnote{The \texttt{make\_co\_wd} and \texttt{make\_o\_ne\_wd} test suites simplify some stages of stellar evolution for numerical convenience.}
The CO model was evolved similarly using the \texttt{make\_co\_wd} test suite, producing a $1.06\,\msun$ white dwarf which we then rescaled to $1.14\,\msun$ using the \texttt{relax\_mass\_scale} procedure (preserving the composition as a function of the relative mass coordinate).
We used the larger \texttt{sagb\_NeNa\_MgAl.net} nuclear reaction network instead of the standard \texttt{co\_burn\_plus.net}, in order to produce more accurate abundances, specifically the oxygen to neon ratio \citep[see][]{DeGeronimo2022}.
We replaced the upper helium atmosphere of the ONe model with hydrogen using \texttt{replace\_initial\_element}, such that both models have similar hydrogen atmospheres, consistent with \phonenumber \citep{DeGeronimo2025}.

We cooled our CO and ONe UMWDs using the test suite \texttt{wd\_cool\_0.6M} down to an effective temperature $T_{\rm eff}=12.4\,\textrm{kK}$, consistent with the measured effective temperature of \phonenumber\,\citep{Jewett2024}.
During the cooling process, the white dwarfs partially crystallize and separate into a solid (crystal) core and a liquid mantle.
In particular, the solid phase is preferentially enriched in the heavier element (oxygen in CO mixtures and neon in ONe mixtures) compared to the liquid.
Conversely, the liquid phase is preferentially enriched in the lighter element (carbon in CO mixtures and oxygen in ONe mixtures) and is less dense than the ambient mixture above it.
The liquid is therefore buoyant and drives compositional convection above the crystal core \citep{Stevenson1980,Mochkovitch1983,Isern1997}.
These effects are computed using the \texttt{phase\_separation} scheme of \citet{Bauer2023}, which utilizes the Skye equation of state \citep{Jermyn2021} and the \citet{BlouinDaligault2021CO,BlouinDaligault2021ONe} phase diagrams.
The numerical scheme mimics convection by mixing the liquid up to an outer convection radius $\rout$, such that the Ledoux stability criterion is satisfied.

As originally suggested by \citet{Isern2017}, these convective motions may power a dynamo and generate strong magnetic fields.
More recently, \citet{CastroTapia2024a} and \citet{Fuentes2024} showed that the magnetic dynamo is efficient only at the very onset of crystallization, when the P\'eclet number is large.
This implies that magnetic field generation is limited to an outer radius $\rout^0$ set by the convective boundary at the onset of crystallization. 
\citet{Fuentes2024} demonstrated that the evolution of $\rout$ at the earliest stages of crystallization is sensitive to modelling details, before stabilizing at a more robust value determined by the white dwarf's composition gradient.
This has led \citet{CastroTapia2024b} to consider $\rout^0$ as a free parameter, resulting in very large uncertainties in the magnetic field. 

\begin{figure*}
        \centering
        \begin{subfigure}[b]{0.5\textwidth}
            \centering
            \includegraphics[width=\textwidth]{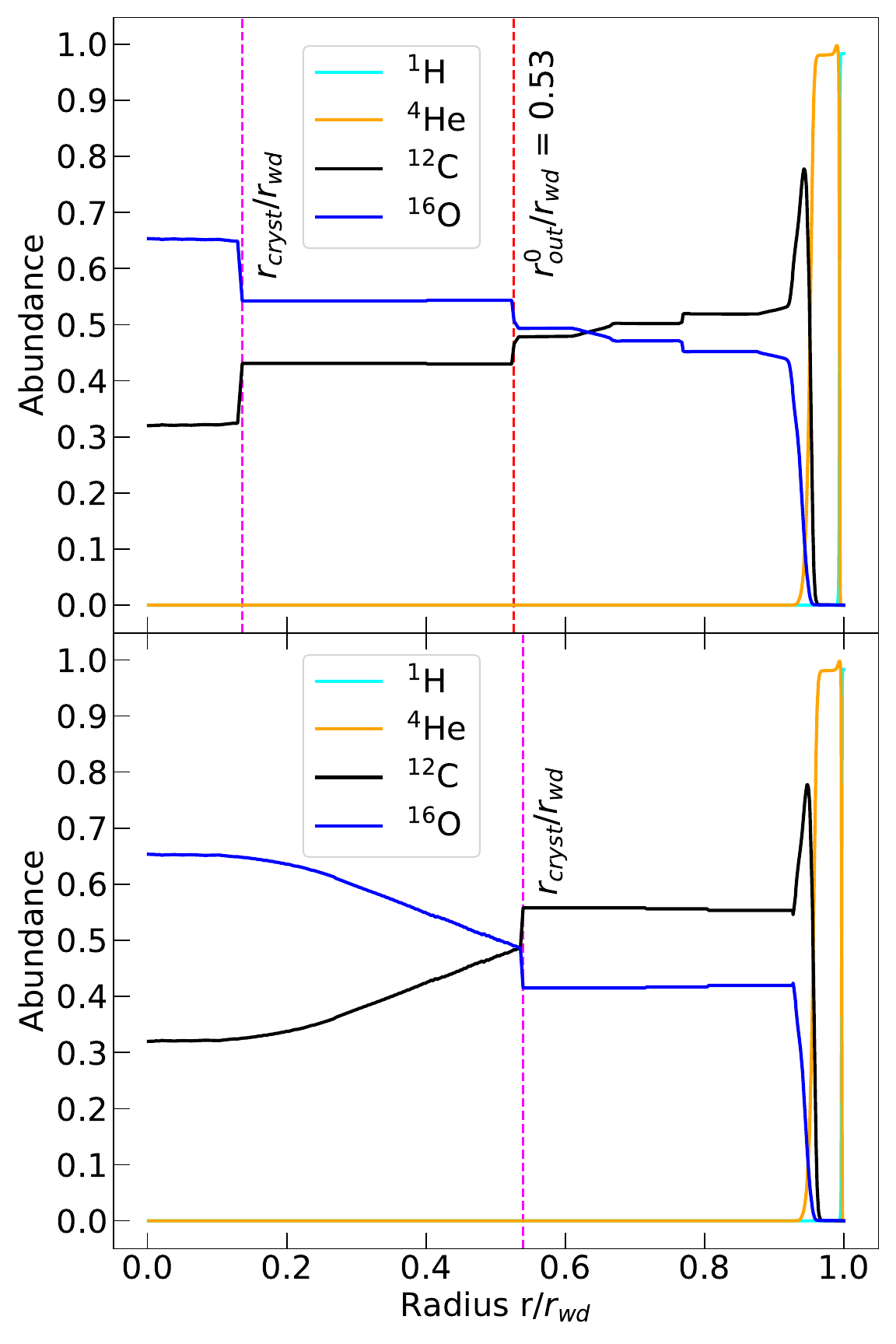}
            \label{fig:Rout_CO}
        \end{subfigure}%
        \hfill
        \begin{subfigure}[b]{0.5\textwidth}   
            \centering 
            \includegraphics[width=\textwidth]{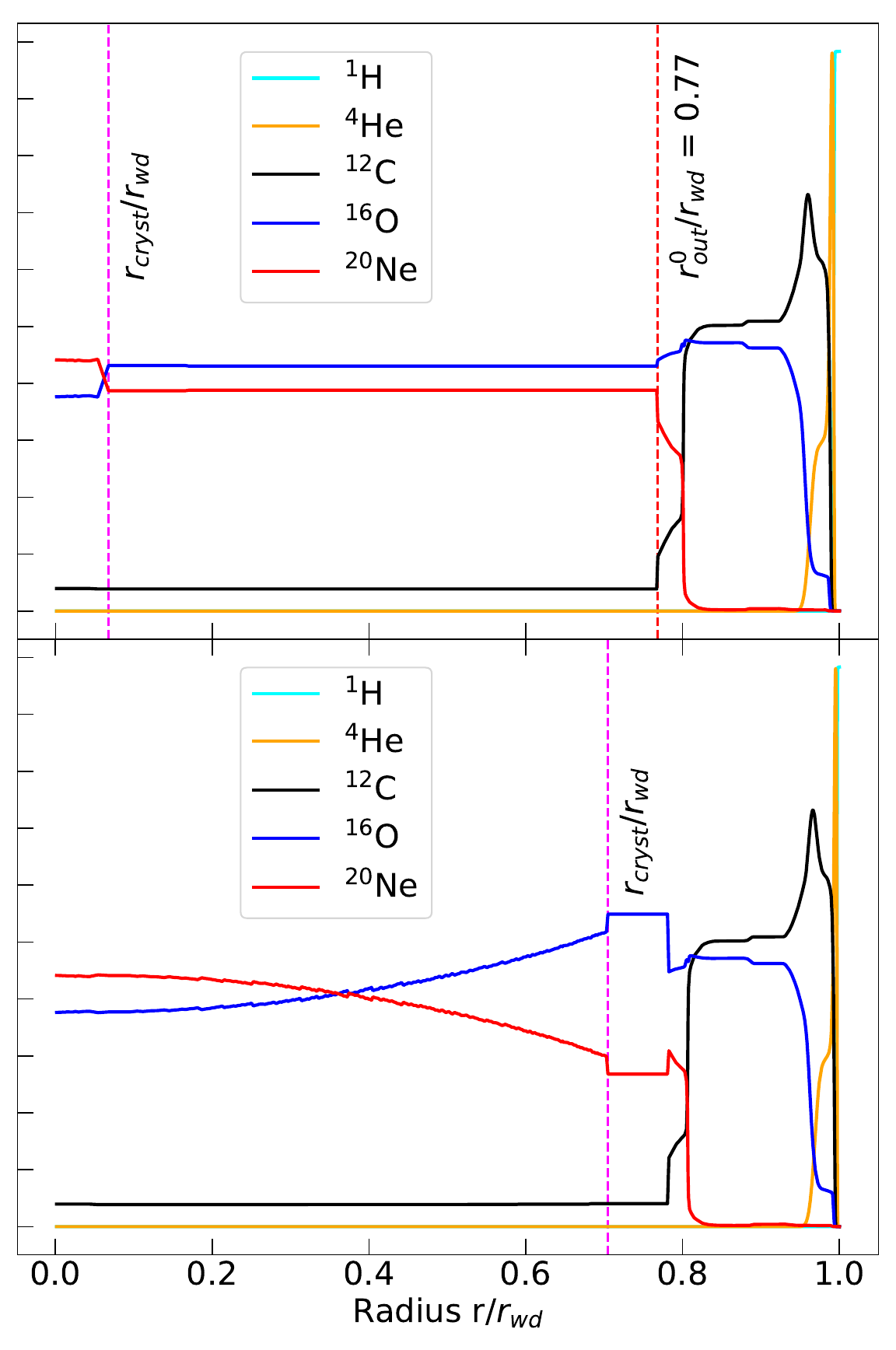}
            \label{fig:Rout_ONe}
        \end{subfigure}%
        \caption{\textit{Top panels:} the composition, crystallization radius $r_{\rm cryst}$, and outer radius of the convection zone $\rout^0$ for a $1.14\,\msun$ white dwarf at the early phase of crystallization -- when a magnetic dynamo is efficient. $\rout^0$ corresponds to the drop in the mean molecular weight in the initial white dwarf profile, which stabilizes convection (see Section \ref{sec:models}). We compute the diffusion of the magnetic field from this $\rout^0$ to the white dwarf's surface $\rwd$ in Section \ref{sec:diffusion}. \textit{Bottom panels:} the white dwarf's structure at a later time, when $T_{\rm eff}=12.4\,\textrm{kK}$, as observed for \phonenumber. \textit{Left:} CO composition. \textit{Right:} ONe composition.}\label{fig:composition}
    \end{figure*}

Here we take a different approach in order to place quantitative bounds on the magnetic field of \phonenumber.
We choose $\rout^0$ shortly after the onset of crystallization, when $\rout$ stabilizes (up to numerical oscillations). As seen in \citet{Fuentes2024}, this time is similar to -- but not exactly -- the duration of efficient convection.\footnote{This is generally true. As seen in fig. 1 of \citet{Fuentes2024}, the saturation of $\rout$ is related to the thermal term in the Ledoux criterion, which also determines whether or not convection is efficient.} Moreover, as seen in their fig. 1, this choice places a robust upper limit on $\rout^0$, insensitive to the details of convection.
As seen in our Fig. \ref{fig:composition}, our models provide $\rout^0/\rwd=0.53$ and $\rout^0/\rwd=0.77$ for the CO and ONe compositions, respectively (where $\rwd$ is the white dwarf's radius).
This outer boundary corresponds to the sharp drop of the mean molecular weight (and the mean atomic mass number) in the initial profile, which stabilizes the compositional convection \citep{BlatmanGinzburg2024}; see \citet{Camisassa2022} for a similar difference in $\rout^0$ between UMWDs of CO and ONe composition.

\section{Magnetic field diffusion}\label{sec:diffusion}

As the magnetic field is only generated in regions unstable to convection and because $\rout<\rwd$, there is a non-negligible delay time between the start of crystallization and the first appearance of magnetic fields at the surface of white dwarfs. Most previous studies estimated this delay by defining an approximate magnetic diffusion time between $\rout$ and the surface \citep{Ginzburg2022,BlatmanGinzburg2024,BlatmanGinzburg2024b,Camisassa2024}. However, diffusion is not a step function -- an exponential tail reaches the surface well before the bulk magnetic field breaks out. To account for this, \citet{CastroTapia2024b} directly solved the induction equation and computed the magnetic field $B(r,t)$ as a function of radius $r$ and time $t$ for $0.5-1.0\,\msun$ CO white dwarfs. Here, we extend their analysis to the ultramassive \phonenumber, assuming both CO and ONe compositions. We briefly repeat the main ingredients of the calculation below for completeness.

We numerically solve the induction equation \citep[e.g.][]{Wendell1987,Cumming2002}:
\begin{equation}
    \label{eq:diffusionEquation}
    \frac{\partial \boldsymbol{B}}{\partial t}= -\boldsymbol{\nabla} \times \left(\eta \boldsymbol{\nabla} \times \boldsymbol{B}\right), 
\end{equation}
where $\eta(r,t)=c^2/(4\upi \sigma)$ is the magnetic diffusivity, $c$ is the speed of light, and $\sigma(r,t)$ is the electric conductivity. The magnetic diffusivity is calculated as in \citet{Cantiello2016}, by interpolating between the non-degenerate, partially degenerate, and fully degenerate regimes \citep{Spitzer1962,Nandkumar1984,Wendell1987}.
We compute $\eta(r,t)$ at any given location and time by interpolating between the \textsc{mesa} models described in Section \ref{sec:models}. Unlike 
\citet{CastroTapia2024b}, who added turbulent diffusivity as a free parameter, we assume that Ohmic diffusion is the only transport mechanism of magnetic fields for $r>\rout^0$ (a higher diffusivity would lower our bounds on the internal field).
We also disregard the effect of the crystal phase on the electric conductivity.
While \citet{CastroTapia2024b} find that this effect could freeze the diffusion of the magnetic field, in our case the crystallization front does not reach the outer layers of the white dwarf (i.e. much beyond $\rout^0$; see Fig. \ref{fig:composition}) and a higher electrical conductivity of the crystal core will likely make little difference.

Assuming an axisymmetric poloidal field with a dominant dipole ($\ell=1$) component, we can define a vector potential $\boldsymbol{A}=A_\phi(r,\theta,t)\hat{\phi}=[R_1(r,t)/r]\sin(\theta) \hat{\phi}$ \citep[see][for details]{Wendell1987,Cumming2002,CastroTapia2024b}.
The magnetic field is then given by
\begin{equation}\label{eq:B}
\boldsymbol{B}=\boldsymbol{\nabla}\times\boldsymbol{A}=B_r\hat{r}+B_\theta\hat{\theta}=\frac{2R_1\cos\theta}{r^2}\hat{r}-\frac{\sin\theta}{r}\frac{\partial R_1}{\partial r}\hat{\theta},
\end{equation}
reducing equation \eqref{eq:diffusionEquation} to
\begin{equation}
    \label{eq:radialEquation}
    \frac{\partial R_1}{\partial t}= \eta(r,t) \left(\frac{\partial ^2 R_1}{\partial r^2} -\frac{2 R_1}{r^2}\right). 
\end{equation}
We solve equation \eqref{eq:radialEquation} numerically using the Crank--Nicolson method.
Demanding a finite solution at the centre and infinity, as well as a vanishing current density $\boldsymbol{J}\propto\boldsymbol{\nabla}\times\boldsymbol{B}$ at the surface, yields the following boundary conditions \citep{Cumming2002}:
\begin{equation}
    \frac{\partial R_1}{\partial r}=\begin{cases}
       2R_1/r & r\to0 \\
       -R_1/r & r\to \rwd.
    \end{cases}
\end{equation}
Similarly to \citet{CastroTapia2024b}, our initial conditions assume a uniform magnetic field $B_0$ in the convection zone
\begin{equation}
    \label{eq:initialCondition}
    \frac{2 R_1(r,t_0)}{r^2}=
    \begin{cases}
      B_0 & r \leq \rout^0  \\
      0  & r> \rout^0,
    \end{cases}
  \end{equation}
where $t_0$ is the time at which $\rout^0$ is determined in Section \ref{sec:models}.
Somewhat differently from \citet{CastroTapia2024b}, we assume that the crystal core at $t=t_0$ is also magnetized, because it begins growing during the epoch of efficient convection and presumably inherits the strong magnetic field of the convecting liquid as it crystallizes.
Due to the core's relatively small size $r_{\rm cryst}$ during this epoch (compared to $\rout^0$; see the top panels in Fig. \ref{fig:composition}) and the high electrical conductivities in the crystal phase \citep{CastroTapia2024b}, assuming otherwise would not have a significant effect on the magnetic field's diffusion. 

After solving equation \eqref{eq:radialEquation}, we can calculate the magnetic field strength $B(r,t)\equiv|\boldsymbol{B}(r,t)|$ using equation \eqref{eq:B}.
For simplicity, we choose $\theta=0$, for which the field is purely radial
\begin{equation}
    B(r,t)=B_r(r,t)=\frac{2 R_1(r,t)}{r^2}.
\end{equation}

\section{Seismic upper limits on the magnetic field}\label{sec:seismology}

\begin{figure*}
    \centering
    \includegraphics[width=\textwidth]{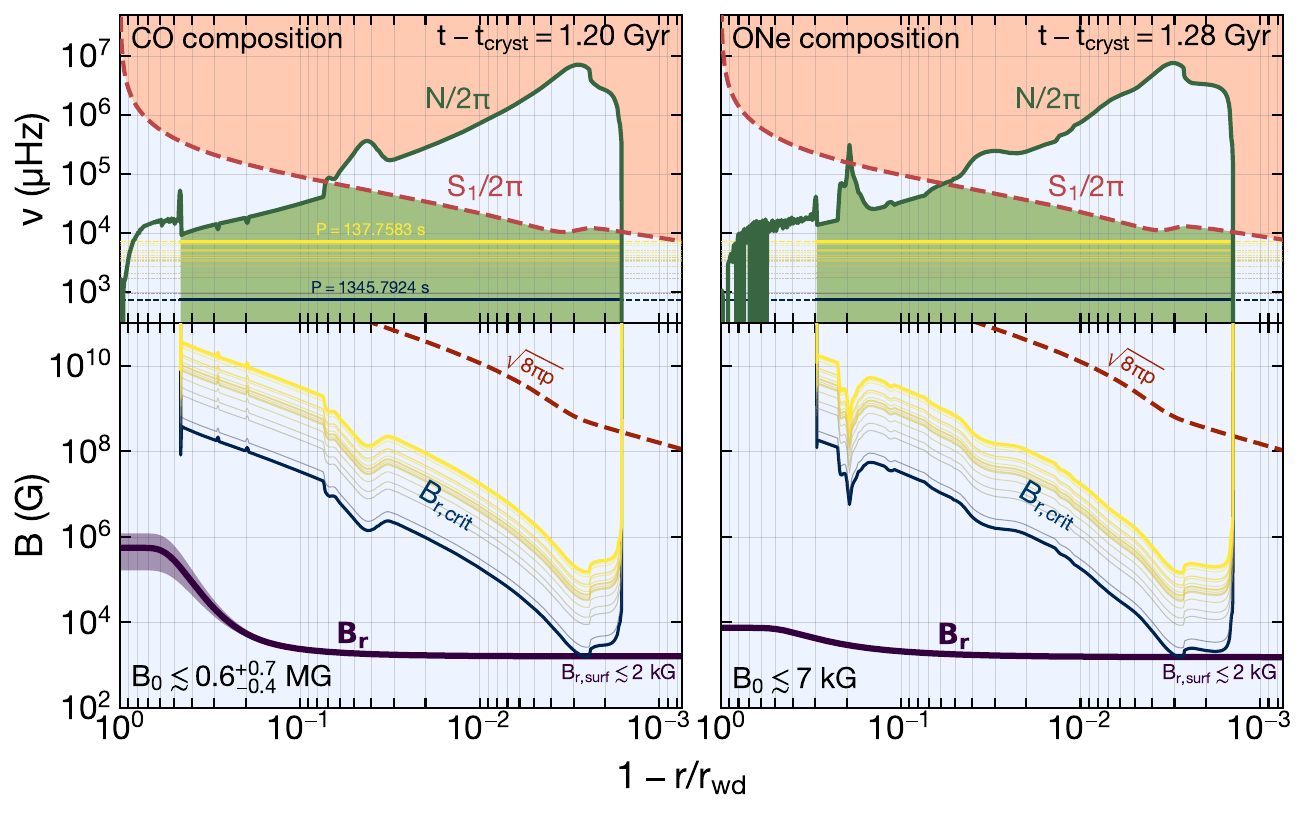}
    \caption{\textit{Top}: asteroseismic propagation diagrams for the CO and ONe white dwarf models when $T_{\rm eff}=12.4\,\mathrm{kK}$.
    The green region denotes the g-mode cavity.
    Observed mode periods of \phonenumber are also indicated.
    \textit{Bottom}: magnetic fields required to suppress the propagation of g modes ($B_{r,\mathrm{crit}}$; solid thin blue-to-yellow lines corresponding to the observed modes) or their convective excitation ($\sqrt{8\upi p}$; brown dashed lines).
    The magnetic field $B_r$ obtained from the diffusion calculation in Section \ref{sec:diffusion} is also shown (thick solid purple lines), normalized to the maximum core field $B_0$ consistent with the observed modes. In the CO white dwarf, only the exponential tail of the field has diffused to the surface, implying a much larger $B_0\gg B_{r,\mathrm{surf}}$, which is sensitive to uncertainties in stellar evolution such as the $^{12}{\rm C}(\alpha,\gamma)^{16}{\rm O}$ nuclear reaction rate (purple shading).}
\label{fig:ultramagnet_seismology}
\end{figure*}

Many white dwarfs pulsate in gravity (g) modes, which propagate in regions where the mode frequency $\omega=2\upi\nu$ is less than both the Brunt--V\"ais\"al\"a frequency $N$ and Lamb frequency $S_\ell$ (the green region in the top panels of Fig. \ref{fig:ultramagnet_seismology}).\footnote{In our case, the inner boundary of the g-mode cavity is set by the crystallization radius $r_{\rm cryst}$, assuming that gravity waves do not penetrate into the solid core \citep{Montgomery1999,Corsico2004,Corsico2005,DeGeronimo2019}.}
The largest class of pulsating white dwarfs is the ZZ Ceti variables, also called DAV white dwarfs.
ZZ Cetis all have hydrogen atmospheres, and are all concentrated in a narrow instability strip centred around $T_{\mathrm{eff}}\simeq11\,\mathrm{kK}$, within which their outer convection zones form and deepen substantially due to partial ionization \citep{FontaineBassard2008,WingetKepler2008}.

Although most ZZ Cetis are typical-mass white dwarfs ($\sim0.6\,\msun$), a few ultramassive ($\gtrsim1.1\,\msun$) pulsating white dwarfs are known \citep{Metcalfe2004,Hermes2013,Curd2017,GentileFusillo2018,Rowan2019,Vincent2020,Kilic2023,Caliskan2025,DeGeronimo2025}.
Of these, \phonenumber has been observed to possess the richest seismology, with $19$ detected modes with periods ranging from $\approx140\,\mathrm{s}$ to $\approx1350\,\mathrm{s}$.
Given its mass, \phonenumber may possess either a CO or an ONe core -- it is difficult to distinguish which \citep[see][]{DeGeronimo2019}.
The bulk composition of \phonenumber greatly informs both its formation and evolution.
For the purposes of this study, the most relevant difference is that the bulk composition affects the size of the convection region that crystallization induces (see Fig. \ref{fig:composition}), as well as the cooling age at which crystallization occurs (ONe white dwarfs crystallize earlier).  

In this study, we use the seismic non-detection of magnetic fields to place upper bounds on the near-surface magnetic field of \phonenumber, assuming both CO and ONe core compositions.
This analysis largely follows that of \citet{rui2025supersensitive}, who place similar constraints (typically $B_r\lesssim10^3$--$10^4\,\mathrm{G}$) on the magnetic fields of some typical-mass CO white dwarfs ($\approx0.6\,\msun$), as well as a low-mass helium-core white dwarf.
If a strong magnetic field is an inevitable consequence of crystallization-induced mixing (which may facilitate field generation or transport), an upper bound on the surface field of a partially crystallized UMWD -- in which the magnetic field is currently diffusing outwards -- may greatly constrain its internal composition.

When magnetic effects become comparable to buoyant forces, g modes are expected to be suppressed \citep{Fuller2015,lecoanet2017conversion}.
This suppression is expected to dramatically decrease the amplitudes of modes and broaden their spectral widths, and it has likely been observed in red giants \citep[][but see \citealt{mosser2017dipole}]{Fuller2015,Cantiello2016,stello2016prevalence}.
For a mode with period $P=2\upi/\omega$, this suppression occurs when the radial component of the magnetic field reaches a critical value
\begin{equation} \label{bcrit}
    B_r > B_{r,\mathrm{crit}} = 8\upi^{5/2}a_{\mathrm{crit}}\frac{\sqrt{\rho}r}{NP^2}\mathrm{,}
\end{equation}
indicated by the blue-to-yellow curves in Fig. \ref{fig:ultramagnet_seismology}.
\citet{Fuller2015} estimate the order-unity prefactor $a_{\mathrm{crit}}$ to be
\begin{equation}\label{eq:acrit}
    a_{\mathrm{crit}} \simeq \frac{1}{2\sqrt{\ell(\ell+1)}} \mathrm{.}
\end{equation}

We hereafter adopt this value of $a_{\mathrm{crit}}$, although it generally depends on the magnetic field geometry and mode numbers in a complicated way \citep{lecoanet2017conversion,rui2023gravity}.
The uncertainty in this prefactor means that seismic upper bounds on the magnetic field are only precise at the order-of-magnitude level.
As in \citet{rui2025supersensitive}, we assume that the detection of a sharply peaked mode implies the absence of magnetic g-mode suppression and bounds $B_r<B_{r,\mathrm{crit}}$ calculated according to that mode's frequency \citep[however, in reality this suppression may not be total, see][]{mosser2017dipole,loi2020magneto,muller2025oscillations}.
We conservatively assume that all peaks detected in \phonenumber correspond to $\ell=1$, which according to equations \eqref{bcrit} and \eqref{eq:acrit} is the least sensitive to magnetism, so $\ell \geq 2$ modes would require even weaker fields than the values calculated below.

Equation \eqref{bcrit} implies that the threshold field strength for suppression is lowest for low-frequency gravity waves propagating through regions of low density $\rho$ and high stratification (i.e. high $N$).
Accordingly, the strongest seismic upper bound on the field strength will be placed by the longest-period mode with $P\approx1350\,\mathrm{s}$.
Moreover, seismology is most sensitive to near-surface fields (bottom panels of Fig. \ref{fig:ultramagnet_seismology}), and accordingly is most constraining when the magnetic field has had time to diffuse to the surface.
Reassuringly, \citet{rui2025supersensitive} find that the sensitivity of pulsations to the magnetic field is largely insensitive to the hydrogen shell mass, as long as it is enough to produce the associated near-surface convection zone.

For both our CO and ONe white dwarfs, the non-suppression of the longest-period observed mode implies that the radial component of the near-surface magnetic field is less than $B_{r,\mathrm{surf}}\lesssim2\,\mathrm{kG}$ (bottom panels of Fig. \ref{fig:ultramagnet_seismology}).
However, via our diffusion calculations (Section \ref{sec:diffusion}), this corresponds to very different initial core magnetic fields, $B_0\simeq0.6\,\mathrm{MG}$ for the CO composition but only $B_0\simeq7\,\mathrm{kG}$ for the ONe composition.
This difference is a direct result of the extent of the initial convective mixing region $\rout^0$. This outer radius is larger in the ONe case (Fig. \ref{fig:composition}), such that the internal magnetic field diffuses and breaks out at the surface ($B_{r,\mathrm{surf}}\sim B_0$) by the observed cooling age of \phonenumber. In the CO case, on the other hand, only the exponential tail of the diffusing field has reached the surface by that time ($B_{r,\mathrm{surf}}\ll B_0$).

From these upper bounds on the magnetic field, it appears unlikely that \phonenumber harbours an ONe core, unless crystallization dynamos are inefficient even in the early phase \citep{CastroTapia2024a,Fuentes2024}. Even in this scenario, crystallization-induced mixing could in principle transport fossil magnetic fields to $\rout^0$, which would have diffused to \phonenumber's surface by the present day and suppressed its longest-period g mode. Mixing could be a double-edged sword though, potentially erasing fossil fields if the right conditions are met \citep[e.g.][]{Cantiello2016,Camisassa2024}. 

The major uncertainty of our analysis is the size of the initial convection zone $\rout^0$. One aspect of this uncertainty -- the convective instability criterion and exact time when $\rout^0$ is determined -- has already been discussed in Section \ref{sec:models} and by \citet{Fuentes2024}.
Another source of uncertainty in $\rout^0$ is its dependence on the white dwarf's initial composition profile, as seen in Fig. \ref{fig:composition}.
To quantify this, we generated alternative CO models with the same mass $\mwd$ (even before rescaling) by evolving progenitor stars with different initial masses and modified $^{12}{\rm C}(\alpha,\gamma)^{16}{\rm O}$ nuclear reaction rates within the $\pm 3\sigma$ uncertainty limits \citep{Mehta2022}.
This poorly constrained nuclear reaction converts carbon into oxygen, and is one of the biggest unknowns in stellar evolution theory, and in white dwarf structure in particular \citep{Chidester2022}.
As demonstrated in Fig. \ref{fig:ultramagnet_seismology}, the different white dwarf structure changes $\rout^0$ and the magnetic field's diffusion.
The exponential tail of the diffusing field is very sensitive to changes in $\rout^0$, leading to a factor of a few uncertainty in $B_0$ for a CO composition (bottom left panel in Fig. \ref{fig:ultramagnet_seismology}; note that $B_{r,\mathrm{crit}}$ in the outer layers is insensitive to the $\alpha$-capture reaction, which modifies primarily the white dwarf's interior).
In the ONe case, magnetic diffusion has already proceeded beyond the exponential phase, and the sensitivity to $\rout^0$ is much smaller ($\Delta B_0/B_0\sim 0.1$, not plotted in Fig. \ref{fig:ultramagnet_seismology}).

We also tested the sensitivity to convective overshooting during the core and shell helium burning phases of the white dwarf's progenitor, during which the CO profile is established. 
Specifically, our \textsc{mesa} calculations employ the exponential overshoot scheme of \citet{Herwig2000}; see \citet{Paxton2011} for details. Our nominal overshoot parameters are $f=1.4\times 10^{-2}$ and $f_0=4\times 10^{-3}$, such that the overshoot mixing extends a fraction $\approx f-f_0=1.0\times 10^{-2}$ of the local scale height beyond the convection zone -- of the same order of magnitude as \citet{Herwig2000} and \citet{Chidester2023}. We also tested reduced ($f=f_0=10^{-3}$) and enhanced ($f=2.8\times 10^{-2}$, $f_0=8\times 10^{-3}$) overshoot models for the CO case, while keeping the same $\mwd$ (even before rescaling) by changing the progenitor's initial mass. While overshooting changes the core mass (and hence $\mwd$) for a given progenitor, it only mildly affects the white dwarf's structure for a given $\mwd$ (which is constrained observationally), resulting in almost the same $\rout^0/\rwd\approx 0.53$ and $B_0$.

Very recently, \citet{Castro-TapiaCumming2025} demonstrated that the structure of ONe white dwarfs during crystallization is sensitive to the ternary nature of their phase diagram, which includes minor species such as sodium or magnesium.  
We emphasize that the white dwarf's structure may also be sensitive to its formation history (single stellar evolution or a merger), which should be tested in future work.

We note that the suppression of pulsations considered here is due to the inhibited \textit{propagation} of g modes due to the magnetic field.
This is different from the magnetic field $\sim\sqrt{8\upi p}$ ($p$ is the pressure) required to suppress near-surface convection \citep[the brown dashed lines in the bottom panels of Fig. \ref{fig:ultramagnet_seismology}, e.g.][]{Tremblay2015}, which is thought to be involved in the \textit{excitation} of the oscillations.
In our white dwarf models, the magnetic field required to suppress convection throughout the entire convective zone (where $N^2$ is negative) is $\gtrsim100\,\mathrm{MG}$.
Magnetic suppression of g-mode propagation therefore imposes much tighter upper limits on the magnetic field.

In fact, the surface convection zone -- if not suppressed -- may generate kG magnetic fields through its own dynamo, regardless of an outwardly diffusing internal field \citep{Fontaine1973,Markiel1994,Thomas1995,Tremblay2015}. Whether or not such fields manage to penetrate into the radiative g-mode cavity (and affect the seismology) remains to be checked; see also \citet{Yaakovyan2025} for a discussion of the surface dynamo. 

\subsection{Sensitivity of mode frequencies to magnetic fields}

\begin{figure}
    \centering
    \includegraphics[width=0.48\textwidth]{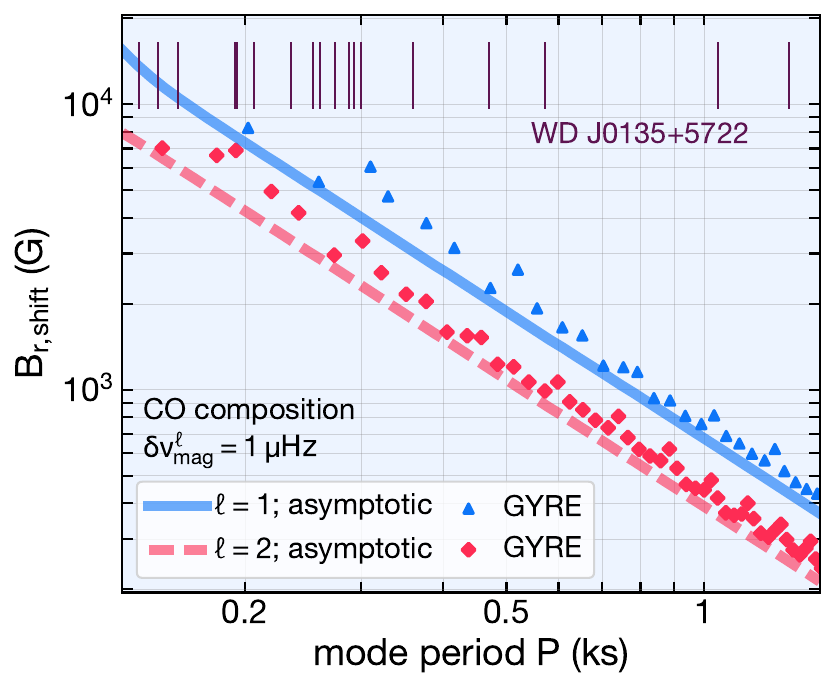}
    \caption{The magnetic field implied by observed magnetic frequency shifts of $\delta\nu_{\mathrm{mag}}^\ell=1\,\mu\mathrm{Hz}$ to the dipole (blue) and quadrupole (red) modes for the CO model.
    Points show estimates calculated using mode eigenfunctions computed using \textsc{gyre}.
    Translucent curves show estimates calculated using equation \eqref{Brseismicshift}, in the asymptotic limit \citep[including the ad hoc non-asymptotic correction factor of][]{rui2025supersensitive}.
    The same curves for the ONe model (not shown here) are very similar.
    Frequencies of oscillation modes observed in \phonenumber are indicated by the vertical lines.
    }
\label{fig:ultramagnet_frequency_shifts}
\end{figure}

This work focuses on the ability of magnetic fields $\gtrsim B_{r,\mathrm{crit}}$ to suppress g-mode oscillations.
However, magnetic fields weaker than $B_{r,\mathrm{crit}}$ can still affect g modes by shifting their frequencies. The frequency shift $\delta \nu_{\rm mag}$ is approximately \citep[see][and references therein, for details]{rui2025supersensitive}
\begin{equation}
    \delta\nu_{\mathrm{mag}} \equiv \frac{\ell(\ell+1)}{128\upi^5}\mathscr{I}\langle B_r^2\rangle P^3\mathrm{,}
\end{equation}
where $\langle B_r^2\rangle$ is the squared magnetic field averaged across the g-mode cavity $\mathcal{R}^\ell_{\;\nu}$ (where $\omega<N,S_\ell$),
\begin{equation} \label{avgfield}
    \langle B_r^2\rangle = \int_{\mathcal{R}^\ell_{\;\nu}}\mathrm{d}r\,K(r)\int \frac{\mathrm{d}\Omega}{4\upi}\,B_r^2 \, \mathrm{.}
\end{equation}
Here, the weight function
\begin{equation} \label{Kasympt}
    K(r) \simeq \begin{cases}
        \frac{N^3/\rho r^3}{\int_{\mathcal{R}^\ell_{\;\nu}}(N^3/\rho r^3)\,\mathrm{d}r} & \mathrm{in}\,\mathcal{R}^\ell_{\;\nu} \\
        0 & \mathrm{otherwise}
    \end{cases}
\end{equation}
peaks sharply near the surface. The sensitivity factor $\mathscr{I}$ is given by
\begin{equation} \label{scriptIasympt}
    \mathscr{I} \simeq \mathscr{I}^{\mathrm{asympt}} \equiv \frac{\int_{\mathcal{R}^\ell_{\;\nu}}(N^3/\rho r^3)\,\mathrm{d}r}{\int_{\mathcal{R}^\ell_{\;\nu}}(N/r)\,\mathrm{d}r}\mathrm{,}
\end{equation}
and depends on the stellar structure.
Because gravity waves are expected to undergo near-total reflection at the solid--liquid interface \citep{Montgomery1999}, we exclude the crystallized region from the g-mode cavity $\mathcal{R}^\ell_{\;\nu}$.
In practice, this makes little difference since it only modifies the inner boundary of the g-mode cavity, where gravity waves are insensitive to magnetism.
The superscript `asympt' denotes that $\mathscr{I}$ was derived under an asymptotic approximation, assuming that the radial wavelength is much smaller than any other relevant length scale.
While $\mathscr{I}^{\mathrm{asympt}}$ generally depends on $\ell$ through the size of the g-mode cavity, both our CO and ONe models imply that the outer boundary of the cavity is set by $N$ for all of the modes observed in \phonenumber, and is therefore approximately independent of $\ell$ (top panels of Fig. \ref{fig:ultramagnet_seismology}).

The observation of a frequency shift $\delta\nu_{\mathrm{mag}}$ of magnetic origin therefore implies a near-surface field
\begin{equation} \label{Brseismicshift}
    B_{r,\mathrm{surf}} \simeq B_{r,\mathrm{shift}} \equiv \sqrt{\frac{128\upi^5}{\ell(\ell+1)}\frac{\delta\nu_{\mathrm{mag}}}{\mathscr{I}P^3}}\mathrm{.}
\end{equation}
Conversely, non-detection of such a frequency shift places a strong seismic upper limit on the magnetic field \citep[see][who place rough upper bounds on $B_{r,\mathrm{surf}}$ for $24$ white dwarfs based on their relatively symmetric rotational splittings]{rui2025supersensitive}.
Using equations \eqref{scriptIasympt} and \eqref{Brseismicshift}, we compute values of $B_{r,\mathrm{shift}}$ required to produce g-mode frequency shifts $\simeq\delta\nu_{\mathrm{mag}}=1\,\mu\mathrm{Hz}$ \citep[comparable to the smallest frequency uncertainties of \phonenumber given by][]{DeGeronimo2025} in dipole and quadrupole modes (translucent curves in Fig. \ref{fig:ultramagnet_frequency_shifts}).
In calculating $B_{r,\mathrm{shift}}$, we include the ad hoc non-asymptotic correction factor described by \citet{rui2025supersensitive}.

Consistent with the explicit scaling $B_{r,\mathrm{shift}}\propto P^{-3/2}$ in equation \eqref{Brseismicshift}, the frequencies of longer-period modes are significantly more sensitive to magnetic fields.
Moreover, in both models, $\mathscr{I}^{\mathrm{asympt}}$ is approximately a constant over the entire frequency range shown in Fig. \ref{fig:ultramagnet_frequency_shifts}.
For the CO and ONe models for a period $P=1350\,\mathrm{s}$, $\mathscr{I}^{\mathrm{asympt}}\approx4.3\times10^{-11}\,\mu\mathrm{Hz}\,\mathrm{G}^{-2}\,\mathrm{s}^{-3}$ and $5.9\times10^{-11}\,\mu\mathrm{Hz}\,\mathrm{G}^{-2}\,\mathrm{s}^{-3}$, respectively.
Unfortunately, neither the unperturbed frequencies nor the quantum numbers of the modes observed in \phonenumber are currently known, and we cannot place an upper limit on $B_{r,\mathrm{surf}}$ based on non-detection of magnetic frequency shifts.
However, we find that a future upper bound on $\delta\nu_{\mathrm{mag}}\lesssim1\,\mu\mathrm{Hz}$ (e.g. due to a relatively symmetric rotational multiplet) for the longest-period mode observed in \phonenumber would translate to a very stringent upper bound on the magnetic field $B_{r,\mathrm{surf}}\lesssim\mathrm{few}\times10^2\,\mathrm{G}$ for both $\ell=1$ and $2$ and for both CO and ONe compositions.

In practice, white dwarf pulsations are often localized to near-surface layers where non-asymptotic effects are important \citep{Brassard1992}.
Following \citet{rui2025supersensitive}, we estimate the magnitude of non-asymptotic effects on $\mathscr{I}$ by computing
\begin{equation} \label{scriptInonasympt}
    \mathscr{I}^{\mathrm{GYRE}} = \frac{\omega^2}{\ell(\ell+1)}\frac{\int_{\mathcal{R}^\ell_{\;\nu}}[\partial_r(r\xi_h)]^2\mathrm{d}r}{\int_{\mathcal{R}^\ell_{\;\nu}}\xi_h^2\rho r^2\,\mathrm{d}r}\mathrm{,}
\end{equation}
which reduces to equation \eqref{scriptIasympt} in the asymptotic limit.
The horizontal-displacement eigenfunctions $\xi_h$ are calculated using \textsc{gyre} 8.0 \citep{Townsend2013} for adiabatic oscillations.
Following \citet{Montgomery1999}, we impose a `hard sphere' boundary condition at the crystallization front, setting the radial displacement $\xi_r=0$ there.
For $\delta\nu_{\mathrm{mag}}^\ell=1\,\mu\mathrm{Hz}$, $B_{r,\mathrm{shift}}$ calculated using $\mathscr{I}^{\mathrm{GYRE}}$ instead of $\mathscr{I}^{\mathrm{asympt}}$ are shown as points in Fig. \ref{fig:ultramagnet_frequency_shifts}.
We find reasonable consistency between the asymptotic and non-asymptotic estimates of $B_{r,\mathrm{shift}}$.

\section{Conclusions}\label{sec:conclusions}

Ultramassive white dwarfs (UMWDs) are prime targets for seismology, because their passage through the ZZ Ceti instability strip coincides with the crystallization of their cores.
Specifically, the difference between CO and ONe crystallization times (and its resulting imprint on the white dwarf's pulsation spectrum) may be used to decipher the core's composition \citep[e.g.][]{Corsico2004,Corsico2005,Kanaan2005,Corsico2019review,Corsico2019,DeGeronimo2019,Althaus2021,Kilic2023}. Here, we proposed a different -- indirect -- technique to probe UMWD interiors with seismology by constraining their magnetic fields.
We applied this technique to the richest pulsating UMWD to date, \phonenumber \citep{DeGeronimo2025}.

We found that a relatively weak near-surface magnetic field $B_{r,\mathrm{surf}}\simeq 2\,\textrm{kG}$ is sufficient to inhibit the propagation of gravity waves in the outer layers of the white dwarf.
The longest-period g modes detected in \phonenumber (which are the most sensitive to the magnetic field) serve as sensitive magnetometers which place a stringent upper limit on $B_{r,\mathrm{surf}}$.
At the same time, crystallization drives convection which may generate strong dynamo fields or advect strong fossil fields to the surface, up to an outer radius $\rout<r_{\rm wd}$.
Specifically, convection is efficient at the early stages of crystallization, when the P\'eclet number $\mathrm{Pe}\gtrsim 1$, and the outer radius $\rout^0$ is set by the white dwarf's initial compositional gradient \citep{CastroTapia2024a,Fuentes2024}.
We computed the Ohmic diffusion of magnetic fields from $\rout^0$ to the surface and translated our seismic limit on $B_{r,\mathrm{surf}}$ to a limit on the internal field $B_0$. 

For a CO core, our method yields $B_0\lesssim 0.6\,\textrm{MG}$, consistent with theoretical estimates of an efficient crystallization dynamo \citep{Ginzburg2022,Fuentes2024} or a fossil field \citep{Camisassa2024}, as well as with observations of late-emerging fields in lower-mass white dwarfs \citep{BagnuloLandstreet2022}.
In the case of \phonenumber, only the exponential tail of the diffusing field has reached the surface by the time it is observed, such that $B_{r,\mathrm{surf}}\ll B_0$.
For an ONe core, on the other hand, $\rout^0$ is larger \citep{Camisassa2022}, such that the bulk field breaks out to the surface and $B_{r,\mathrm{surf}}\lesssim B_0\lesssim 7\,\textrm{kG}$. This exceedingly low upper bound on the internal field $B_0$ rules out a short intense dynamo at the onset of crystallization \citep{Fuentes2024}, as well as magnetization by a merger \citep{GarciaBerro2012}; alternatively, it rules out an ONe core composition for \phonenumber. 

We tested the robustness of our arguments to some details of stellar evolution by modifying the UMWD progenitor's $^{12}{\rm C}(\alpha,\gamma)^{16}{\rm O}$ nuclear reaction rate (which changes $\rout^0$) and overshoot parameters, but other uncertainties remain.
For example, the UMWD structure and $\rout^0$ may be sensitive to its formation path (single stellar evolution or a double white dwarf merger).
During the white dwarf cooling phase itself, the compositional mixing associated with crystallization is not yet fully understood \citep[specifically the location of $\rout^0$; see][]{Fuentes2024}, and so is the transport of magnetic fields through the layers above $\rout^0$ \citep[e.g. turbulent diffusivity; see][]{CastroTapia2024b}.

Never the less, seismology provides stringent upper limits on UMWD surface magnetic fields, which may become even tighter if magnetic shifts in the pulsation frequency can be constrained (our current limits for \phonenumber rely merely on the detection of pulsation modes, regardless of any shift in their frequency). As demonstrated here, these tight upper limits can constrain the exponential tail of diffusing internal fields that are still trapped inside the white dwarf's interior. Thus, under certain assumptions, seismic magnetometry -- which is sensitive to near-surface magnetic fields -- may be used to probe UMWD interiors, and help reveal the formation channels and core compositions of this intriguing population.  

\section*{Acknowledgements}

We thank Maria Camisassa, JJ Hermes, and Eliot Quataert for helpful discussions.
We are grateful for support from the United States-Israel Binational Science Foundation (BSF; grant no. 2022175).
DB and SG are also supported by the Israel Ministry of Innovation, Science, and Technology (grant no. 1001572596), the Israel Science Foundation (ISF; grant nos 1600/24 and 1965/24), and the German-Israeli Foundation for Scientific Research and Development (GIF; grant no. I-1567-303.5-2024).
NZR acknowledges support from the National Science Foundation Graduate Research Fellowship under grant no. DGE‐1745301.

\section*{Data Availability}

The data underlying this article will be shared on reasonable request to the corresponding authors.



\bibliographystyle{mnras}
\input{seismic.bbl}


\bsp	
\label{lastpage}
\end{document}